\documentclass[prb,twocolumn,superscriptaddress]{revtex4}
\usepackage{amsmath,amssymb,graphicx,color,bm,epstopdf}

\begin{document}
\title{BCS pairing in fully repulsive fermion mixtures}

\author{T. Espinosa-Ortega}
\affiliation{Division of Physics and Applied Physics, Nanyang Technological University 637371, Singapore}

\author{O. Kyriienko}
\affiliation{Division of Physics and Applied Physics, Nanyang Technological University 637371, Singapore}
\affiliation{Science Institute, University of Iceland, Dunhagi-3,
IS-107, Reykjavik, Iceland}

\author{I. A. Shelykh}
\affiliation{Division of Physics and Applied Physics, Nanyang Technological University 637371, Singapore}
\affiliation{Science Institute, University of Iceland, Dunhagi-3,
IS-107, Reykjavik, Iceland}

\date{\today}

\begin{abstract}
We consider a mixture of two neutral cold Fermi gases with repulsive interactions. We show that in some region of the parameter space of the system the effective attraction between fermions of the same type can appear due to the exchange of collective excitations. This leads to the formation of BCS pairing in the case where bare inter-atomic interactions are repulsive.
\end{abstract}


\maketitle

%*******************************************************************************************************

\section{Introduction} The physics of ultra cold atoms has attracted much attention in recent decades. One of the reasons is the wide variety of quantum collective phenomena which can be experimentally observed in these systems. Besides, cold atoms can be used as an outstanding playground for investigation of condensed matter analogies of phenomena belonging to the domain of cosmology, allowing to simulate the behavior of such objects as neutron stars and black holes.\cite{Feynman,Weimer,Gezerlis}

One of the main challenges in atomic physics was the observation of a Bose-Einstein condensate (BEC) -- a macroscopically occupied state formed by bosonic atoms cooled down beyond the critical temperature --\cite{ReviewBEC} which was achieved for the first time in 1995.\cite{Anderson,Cornell,Davis,Ketterle} This discovery stimulated activity in the field of cold bosons and led to the experimental investigation of such phenomena as: quantized vortices and vortex lattices;\cite{Matthews} solitons;\cite{Ostrovskaya} optical trapping;\cite{Bloch} the Josephson effect and self trapping;\cite{Albiez} the BKT transition\cite{Hadzibabic} and others.

On the other hand, it was realized that cold fermions can also reveal interesting physics. Major attention was attracted to the BEC-BCS crossover driven by the phenomenon known as Feshbach resonance.\cite{Courteille} It originates from the tuning of atomic multiplets with magnetic field and corresponding change of sign of the scattering length for atoms at the critical magnetic field $B_{0}$.\cite{FeshbachReview} Below the Feshbach resonance, the scattering length is positive but individual fermionic atoms are bound into molecules, which have bosonic properties and at low temperature form BEC.\cite{Regal} On the other hand, for $B>B_{0}$ the molecules unbound, but the scattering length becomes negative. In this regime the analog of a Bardeen-Cooper-Schrieffer (BCS) many-body state is formed.\cite{BCS} The latter was first proposed for the description of the phenomenon of superconductivity which occurs in certain metals and alloys and is characterized by formation of loosely bound Cooper pairs. Finally, in the vicinity of critical magnetic field, $B_{0}$, the scattering length diverges which leads to strong correlations in the system and onset of the universal behavior.\cite{Zwerger}

It was recently supposed that the physics of the BEC-BCS crossover can be important for understanding of mechanisms of unconventional high-temperature superconductivity (HTS). The phenomenon was detected experimentally for a copper compound,\cite{Bednorz} but the corresponding theory is still lacking.\cite{Dagotto,Legget} This stimulated proposals of novel methods of BCS state formation, \textit{e.g.}, for the hybrid electron-polariton system,\cite{LaussyPRL} where the effective attractive interaction emerges due to electron-exciton interactions.

In this article we present the idea of a new possible mechanism of BCS pairing in a cold fermion system. We study the two-component Fermi mixture of neutral atoms where all bare particle-particle interactions are repulsive. However, the total effective inter-particle interaction should account for the possibility of virtual collective excitations analogous to plasmons in solid state systems. In a gas of identical fermions this leads to the screening of the inter-particle repulsion. Here we show that in a binary mixture similar corrections can lead to the onset of an effective attraction and subsequent BCS state formation. We analyze the required parameters for observation of this phenomenon and present the corresponding phase diagrams.

\section{The model} We consider the ground state of  binary fermionic mixture at ultra-low temperature. The particle-particle interactions are treated using the s-scattering approximation, which is applicable if particles are electrically neutral and do not have a dipole moment. In this case the bare interactions in the system are fully characterized by three scattering lengths $a_{11}$, $a_{22}$ and $a_{12}$, corresponding to: the interactions between two atoms of type $1$; two atoms of type $2$; and between one atom of type $1$ and one atom of type $2$, respectively.

The matrix elements of the bare interaction between particles of the same kind are given by\cite{Bloch}
\begin{equation}
\label{Vbare}
V_{11,22}=\frac{4 \pi a_{11,22}}{m_{1,2}}
\end{equation}
and for interaction between different particles
\begin{equation}
V_{12}=\frac{2 \pi a_{12}}{\mu}
\end{equation}
where $m_{1,2}$ are masses of the particles and $\mu=m_1 m_2/(m_1 + m_2)$ denotes the reduced mass. Here we assume all scattering lengths to be positive which means that all interactions are repulsive.

Our goal is to calculate the effective interactions between atoms of the same type accounting for the possibility of creation of virtual collective excitations in the ground state of the system. The convenient way to do this is to use a language of Feynman diagrams. While the general task of accounting for all possible diagrams can not be achieved, there are several approximations which allow to explain certain aspects of Fermi gas behavior in particular limiting cases. Going beyond the mean-field theory, the diagrammatic series, which includes interparticle interaction, can be often written using the so called random phase approximation (RPA).\cite{Pines,Koch} The important consequence of the RPA is a correct description of the bulk plasmons -- collective oscillations of the density. In the context of the electronic plasma in metals, the account of the RPA corrections leads to the screening of the Coulomb interaction, which can be expressed as
\begin{equation}
V_{eff}=\frac{V_{0}(q)}{1-V_{0}(q)\Pi (q,\omega)}=\frac{V_{0}(q)}{\varepsilon(q,\omega)},
\label{Veff1}
\end{equation}
where $V_{0}(q)$ is a bare Coulomb interaction and $\Pi (q,\omega)$ denotes a polarization operator. $\varepsilon(q,\omega)=1-V_{0}(q)\Pi (q,\omega)$ is the dielectric function, which is responsible for charge screening and in general is frequency dependent.
\begin{figure}
\includegraphics[width=0.6\linewidth]{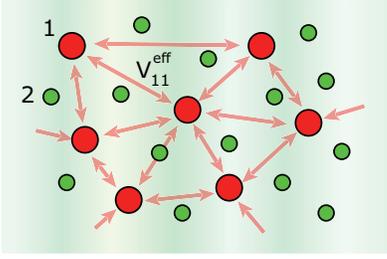}
\caption{(Color online). Sketch of the system representing a two component Fermi mixture with atoms of the first (red circles) and second (green circles) kind. The red arrows schematically depict the effective interaction between particles $V^{eff}_{11}$, which is mediated by the density modulations of the Fermi gas consisting of particles of type $2$.}
\label{Fig1}
\end{figure}

In the case of an electron gas in metals, the RPA is applicable in the regime of high densities. The condition is that the kinetic energy of Fermi gas should overcome its potential energy. The kinetic energy of the three dimensional Fermi gas scales as $E_{kin}\sim n^{5/3}$, where $n$ is the concentration of the gas. The potential energy of the Coulomb interaction scales as $E^{el}_{pot}\sim n^{4/3}$. One sees, that $E_{kin}$ is a faster growing function of $n$, which makes theoretical description of the dense electron plasma easier than description of the diluted electron gas, for which the account of correlation corrections becomes a very tricky task.

However, in the domain of the cold fermion atoms the situation becomes qualitatively different. Differently from the Coulomb case, the interaction between neutral atoms is short range, and the potential energy of the atomic gas scales as the square of its concentration $E_{pot}=V L^3 n^2$, with $V=4\pi a/m$ being the interaction constant, $L^3$ denotes volume and $m$ is the mass of a particle. As function of $n$, it grows faster than the kinetic energy, which means that the RPA description in this case should work better in the low density limit, the condition which is usually satisfied in the experiments. We will therefore use the RPA approximation in our further consideration.

\begin{figure}
\includegraphics[width=1.\linewidth]{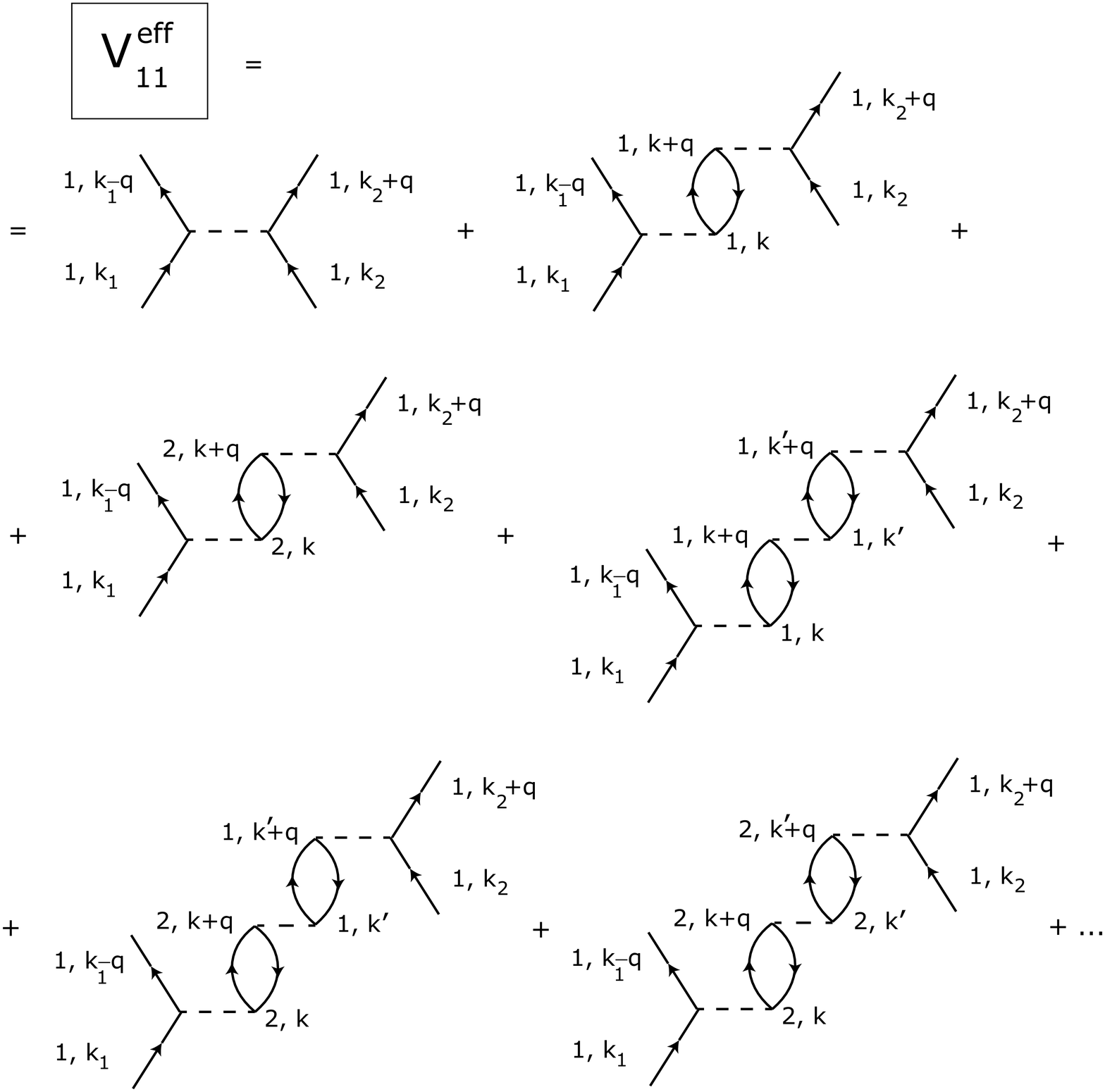}
\caption{Diagrammatic representation of the effective interaction between the atoms of the first kind mediated by the collective excitations in a two-component Fermi gas.}
\label{Fig2}
\end{figure}

In the case of multi-fermion mixtures the effective interaction can be written using the generalized dielectric function\cite{Shelykh,Kyriienko}
\begin{equation}
\mathbf{V}_{eff}=\mathbf{V}_{0}\left(1-\mathbf{V}_{0}\mathbf{\Pi}\right)^{-1},
\label{Veff}
\end{equation}
where by bold characters $\mathbf{V}_0$ and $\mathbf{\Pi}$ we denote $2\times 2$ matrices of the bare interactions and polarization operators describing the system,
\begin{equation}
\mathbf{V}_{0} =
\left( \begin{array}{cc}
V_{11} & V_{12} \\
V_{12} & V_{22} \\
\end{array} \right),
\end{equation}

\begin{equation}
\mathbf{\Pi} =
\left( \begin{array}{cc}
\Pi_{1} & 0\\
0 & \Pi_{2}\\
\end{array} \right).
\end{equation}
Subscripts $1$ and $2$ correspond to the fermion flavor in the system. Expression (\ref{Veff}) can be obtained by performing summation of all RPA type diagrams shown in Fig. \ref{Fig2}.

The polarization operator for the Fermi gas can be calculated using the standard rules of evaluation the bubble diagrams and, for zero temperature, for the 3D case, reads \cite{Bruus}
\begin{equation}
\Pi(q,\omega)=-d(E_{F})\Big(1+\frac{f(x,x_0)+f(x,-x_0)}{4x}\Big)
\label{Pi3D}
\end{equation}
where the function $f(x,x_0)$ is defined as
\begin{equation}
f(x,x_0)=\Big(1-(x_{0}/x-x)^{2}\Big)\ln \Big|\frac{x+x^{2}-x_{0}}{x-x^{2}+x_{0}}\Big|
\end{equation}
where we used dimensionless variables $x=q/2k_{F}$ and $x_{0}=\omega /4E_{F}$, and $d(E)$ denotes the density of states. One should note that the polarization operator (\ref{Pi3D}) changes the sign at a characteristic frequency $\omega_{c}$ dependent on $q$.

\begin{figure}
\includegraphics[width=0.9\linewidth]{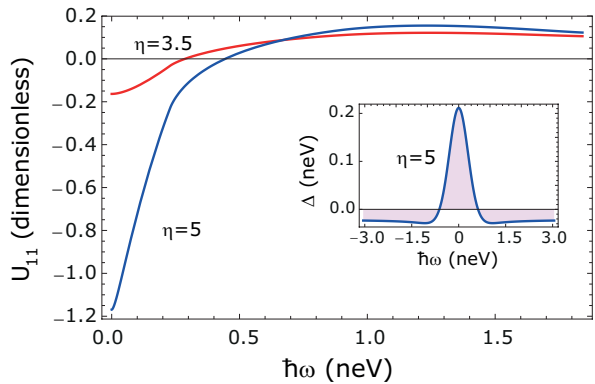}
\caption{(Color online). The model effective averaged potential $U_{11}(\omega)$ plotted for two different ratios $\eta$ ($\eta=3.5$ for red lines and $\eta=5$ for the blue ones). In the inset we plot the gap as a function of frequency for $\eta=5$. The model parameters used for calculations are $a_{11}=a_{22}=200 a_{0}$ where $a_{0}$ denotes the Bohr radius.}
\label{Fig3}
\end{figure}
\begin{figure}
\includegraphics[width=0.9\linewidth]{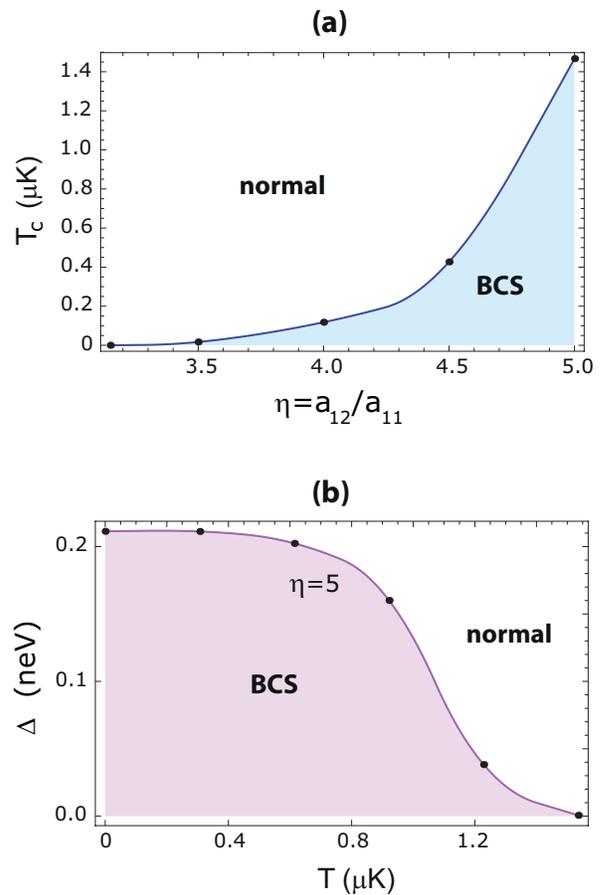}
\caption{(Color online). (a) BCS phase diagram as a function of interaction ratio $\eta=a_{12}/a_{11}$ and the critical temperature $T_{c}$. (b) The temperature dependence of the gap $\Delta$ plotted for $\eta=5$ which allows to find the critical temperature of BCS to normal phase transition.}
\label{Fig4}
\end{figure}

Matrix equation \ref{Veff} allows to obtain the effective interaction between particles of the first kind:
\begin{equation}
V_{11}^{eff}(q,\omega)=\frac{V_{11}(1-\Pi_{2}V_{22})+\Pi_{2}V_{12}^{2}}{(1-\Pi_{1}V_{11})(1-\Pi_{2}V_{22})-\Pi_{1}\Pi_{2}V_{12}^{2}}
\label{V11eff}
\end{equation}

As expected, $V_{11}^{eff}(q,\omega)$ becomes a function of momentum and frequency. The first term in the numerator corresponds to the screened direct interaction between particles of type $1$, while the second term describes the interactions mediated by collective excitation in the system of particles of type $2$ proportional to $V_{12}^2$. Tuning the interaction between particles of different kind, \textit{i. e.}, making $V_{12}\gg V_{11}$, one can make the second term dominant. Therefore, in the low frequency region where $\Pi_{2}$ is negative it is possible to reach the regime where the total effective interaction $V_{11}^{eff}$ becomes attractive. In this situation one can expect that the atoms of the type $1$ will form the BCS state, analogous to those formed by the electrons in a superconductor. The role played by the phonons in a superconductor in our case is played by collective excitations of the system $2$ analogous to bulk plasmons in a metal. The properties of the BCS state, such as the value of the gap and critical temperature, crucially depend on the relative strengths of the interaction between fermions of type $1$ and fermions of types $1$ and $2$, which can be described by the parameter
\begin{equation}
\eta=\frac{a_{12}}{a_{11}}.
\end{equation}

If two fermions interact attractively, even weakly, they can form a Cooper pair. The largest contribution to the process is given by the fermions close to the Fermi surface. Therefore, in the 3D case one can compute an effective dimensionless interaction $U_{11}(\omega)$ between the atoms of type $1$ by averaging the potential of interaction
$V_{11}^{eff}$ over the 2D Fermi sphere, analogous to Ref. [\onlinecite{LaussyJNP}] where the case of 2D polariton-mediated superconductivity was considered. One has:
\begin{equation}
U_{11}(\omega)=\int_{0}^{2\pi}d\phi\int_{0}^{\pi}V_{11}^{eff}(q,\omega)k_{F}^{2}\sin\theta d\theta  /\cal{N},
\label{Veff_omega}
\end{equation}
where $q=\sqrt{2k^2_{F}(1+\cos(\theta))}$ is an exchanged momentum between two fermions lying on a Fermi surface. The normalization factor $\cal{N}$ $=d(E_{F})= m k_{F}/ \pi \hbar^2$ corresponding to the density of states evaluated at the Fermi energy is added to make the effective potential dimensionless.

The shape of $U_{11}(\omega)$ is shown in Fig. \ref{Fig3} for two different values of $\eta$.  One can note the following. First, the effective interaction potential is frequency-dependent, which means that retardation effects play an important role. Second, it crucially depends on $\eta$. For small $\eta$, the direct repulsion between atoms of type $1$ dominates, and $U_{11}(\omega)$ is always positive. The increase of $\eta$ leads to increase of the role played by interactions mediated by collective excitations, and a region of negative values of $U_{11}(\omega)$ develops at small frequencies. This corresponds to the onset of effective attraction at large retardation times. The situation is fully analogous to those in superconducting materials, where retarded phonon-mediated interactions dominate over direct Coulomb repulsion.\cite{Ginzburg} Naturally, the attraction becomes stronger with increase of $\eta$.
\begin{figure}[h]
\includegraphics[width=0.9\linewidth]{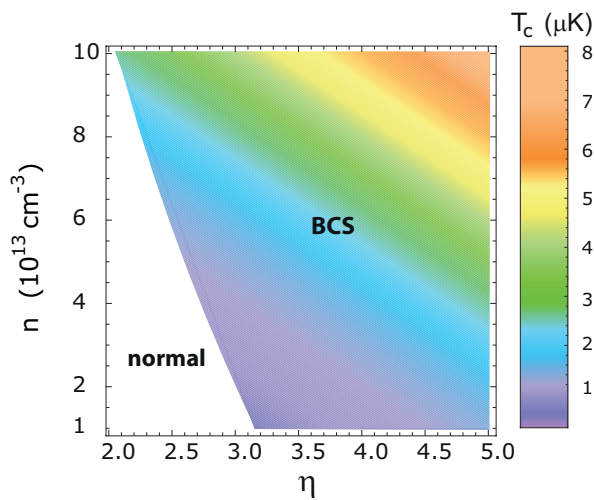}
\caption{(Color online). Phase diagram showing the BCS/normal phase transition as a function of the concentration of particles of type $2$, denoted as $n$, and the interaction ratio $\eta$. The colors indicate the critical temperature in the gapped state.}
\label{Fig5}
\end{figure}

To find the gap in the spectrum connected to emergence of the BCS state and Cooper pair formation, we solve the gap equation in integral form \cite{LaussyPRL,Ginzburg}
\begin{equation}
\Delta (\omega ,T)=-\hbar\int_{-\infty}^{\infty}\frac{U_{11}(\omega -\omega')\Delta(\omega',T)\tanh(E/2k_{B}T)}{2E}d\omega',
\label{gap_eq}
\end{equation}
where: $E=\sqrt{(\hbar\omega')^{2}+E_{0}(k)}$; $E_{0}=\hbar^{2}k^{2}/2 m_1$ is the kinetic energy of atoms of type $1$; and $T$ denotes the temperature.

\section{Results and discussion}

The gap equation is a non-linear integral equation that can be solved numerically by an iterative procedure. The behavior of the gap function is shown in the inset of Fig.\ref{Fig3}. It is if fully analogous to those observed in superconducting solid state systems. The gap is maximal for $\omega=0$, changes sign at some characteristic frequency and then decays to zero when $\omega\rightarrow\infty$.

Figure \ref{Fig4}(a) shows the dependence of the critical temperature $T_c$ on the ratio $\eta$ found by solving Eq. (\ref{gap_eq}). $T_c$ was determined as the value of the temperature above which the gap equation has only a trivial solution, $\Delta=0$, in the whole region of $\omega$.  The concentrations of the fermions of types $1$ and $2$ were chosen as $n_1=n_2=10^{13}$ cm$^{-3}$, and atoms of type $2$ were taken to be four times heavier than atoms of type $1$. There exists a critical value of the parameter $\eta$ below which BCS pairing is absent even at $T=0$ ($\eta_c \approx 3.15$ for the values of the parameters we consider). The increase of $\eta$ above its critical value naturally leads to the increase of the critical temperature.

For a given value of $\eta$ the maximum value of the gap decreases as a function of the temperature and vanishes as $T\rightarrow T_c$, as shown in Fig. \ref{Fig4}(b) for $\eta=5$.

Finally, in Fig. \ref{Fig5} we present the phase diagram in the axes of density of the Fermi gas ($n=n_1=n_2$) and the interaction ratio $\eta$. The concentration of the gas is a relevant parameter since it enters the polarization operator and defines the cut-off energy $\hbar \omega_{c}$ for which the effective interaction potential changes from attractive to repulsive. For the BCS phase we use color to plot the related critical temperature. As it can be expected, the critical temperature of the transition to the BCS state increases with the increase of the concentration and the ratio $\eta$.

\section{Conclusions} We studied the influence of virtual collective excitations represented by diagrams of the RPA type on physical properties of a two-component mixture of neutral fermions with repulsive interactions. We have shown that, in the regime where the interaction between particles of the same type is much weaker than interaction between particles of different types, the total effective interaction can become attractive. This can lead to BCS pairing in the system. The critical temperature of the transition to the BCS state is determined by the concentration and the value of the parameter $\eta$ which characterizes the ratio of the interaction between particles of the same type and particles of different types.

We thank Prof. A. V. Kavokin, Dr. H. Ouerdane and Dr. T.C.H. Liew for valuable discussion. O. Kyriienko acknowledges support from the Eimskip foundation.

\end{document}